\documentclass[letterpaper]{article} 
\usepackage{aaai2026}  
\usepackage{times}  
\usepackage{helvet}  
\usepackage{courier}  
\usepackage[hyphens]{url}  
\usepackage{graphicx} 
\usepackage{natbib}  
\usepackage{caption} 
\usepackage{algorithm}
\usepackage{algorithmic}

\usepackage{newfloat}
\usepackage{listings}
\DeclareCaptionStyle{ruled}{labelfont=normalfont,labelsep=colon,strut=off} 
\floatstyle{ruled}
\newfloat{listing}{tb}{lst}{}
\floatname{listing}{Listing}
\usepackage{xspace}
\usepackage{tikz}
\usetikzlibrary{trees, positioning, shadows, shapes.geometric, calc, arrows, fit}
\pgfdeclarelayer{background}
\pgfsetlayers{background,main}
\usepackage{amsthm}
\theoremstyle{definition}
\usepackage{amssymb} 
\usepackage{booktabs} 

\newcommand{\ac}[1]{\gls{#1}}
\newcommand{\acp}[1]{\glspl{#1}}
\newcommand{\etal}{et al.\xspace}
\newcommand{\etc}{etc.\xspace}
\newcommand{\eg}{e.g.\xspace}
\newcommand{\ie}{i.e.\xspace}
\newcommand{\worksworld}{\textsc{Worksworld}\xspace}

\definecolor{oiBlue}{cmyk}{1, 0.36, 0, 0.30}      
\definecolor{oiSkyBlue}{cmyk}{0.63, 0.23, 0, 0.09}  
\definecolor{oiOrange}{cmyk}{0, 0.31, 1, 0.10}    
\definecolor{oiGray}{cmyk}{0, 0, 0, 0.22}    
\definecolor{tikzYellow}{cmyk}{0, 0, 0.30, 0}       
\definecolor{tikzGreenLight}{cmyk}{0.25, 0, 0.25, 0} 
\definecolor{tikzGreenFaint}{cmyk}{0.08, 0, 0.08, 0} 
\definecolor{tikzGreenDark}{cmyk}{1, 0, 1, 0.5}      

\title{\worksworld{}: A Domain for Integrated Numeric Planning and Scheduling of Distributed Pipelined Workflows}
\author{
    Taylor Paul\textsuperscript{\rm 1},
    William Regli\textsuperscript{\rm 1}
}
\affiliations{
    \textsuperscript{\rm 1}University of Maryland, College Park, MD USA\\
    thpaul@umd.edu, regli@umd.edu
}

\usepackage[acronym]{glossaries}
\setacronymstyle{long-short} 

\newacronym{ai}{AI}{artificial intelligence}
\newacronym{api}{API}{application programming interface}
\newacronym{anns}{ANNS}{approximate nearest neighbor search}
\newacronym{asn}{ASN}{autonomous system number}
\newacronym{aibr}{AIBR}{Additive Interval-Based Relaxation }

\newacronym{cicd}{CI/CD}{continuous integration and continuous delivery}
\newacronym{ci}{CI}{cyberinfratructure}
\newacronym{csp}{CSP}{cloud service provider}

\newacronym{dag}{DAG}{directed acyclic graph}
\newacronym{dds}{DDS}{Data Distribution Service}
\newacronym{dms}{DMS}{data management system}
\newacronym{dpi}{DPI}{data processing interface}
\newacronym{dsi}{DSI}{data sharing interface}

\newacronym{ets}{ETS}{energy-to-solution}
\newacronym{enhsp}{ENHSP}{Expressive Numeric Heuristic Search Planner}

\newacronym{geo}{GEO}{geostationary earth orbit}

\newacronym{hddl}{HDDL}{Hierarchical Domain Definition Language}
\newacronym{hpc}{HPC}{high-performance computing}
\newacronym{htn}{HTN}{hierarchical task network}

\newacronym{iac}{IaC}{Infrastructure as Code}
\newacronym{icn}{ICN}{information-centric networking}
\newacronym{iot}{IoT}{Internet of Things}
\newacronym{ipc}{IPC}{International Planning Conference}

\newacronym{lan}{LAN}{local area network}
\newacronym{leo}{LEO}{low earth orbit}

\newacronym{mrp}{MRP}{multi-repetition relaxed plan}

\newacronym{ndn}{NDN}{named data networking}

\newacronym{osi}{OSI}{Open Systems Interconnection}
\newacronym{oss}{OSS}{open source software}

\newacronym{pddl}{PDDL}{Planning Domain Definition Language}
\newacronym{ps}{Pub/Sub}{publish-subscribe}

\newacronym{qoe}{QoE}{quality of experience}
\newacronym{qos}{QoS}{quality of service}

\newacronym{rbac}{RBAC}{role-based access control}
\newacronym{roi}{ROI}{return on investment}
\newacronym{rtt}{RTT}{round-trip time}

\newacronym{sat}{SAT}{Boolean Satisfiability Problem}
\newacronym{soa}{SOA}{state-of-the-art}
\newacronym{sdn}{SDN}{software-defined networking}
\newacronym{soar}{SOAR}{security orchestration, automation, and response}

\newacronym{tcp}{TCP}{Transmission Control Protocol}
\newacronym{tcpip}{TCP/IP}{Transmission Control Protocol/Internet Protocol}
\newacronym{tts}{TTS}{time-to-solution}

\newacronym{wan}{WAN}{wide-area network}

\newacronym{yaml}{YAML}{YAML Ain't Markup Language}

\begin{document}

\newtheorem{definition}{Definition} 

\maketitle

\begin{abstract}

    This work pursues automated planning and scheduling of distributed data
    pipelines, or workflows. We develop a general workflow and resource graph
    representation that includes both data processing and sharing components
    with corresponding network interfaces for scheduling. Leveraging these
    graphs, we introduce \worksworld{}, a new domain for numeric
    domain-independent planners designed for permanently scheduled workflows,
    like ingest pipelines. Our framework permits users to define data sources,
    available workflow components, and desired data destinations and formats
    without explicitly declaring the entire workflow graph as a goal. The
    planner solves a joint planning and scheduling problem, producing a plan
    that both builds the workflow graph and schedules its components on the
    resource graph. We empirically show that a state-of-the-art numeric planner
    running on commodity hardware with one hour of CPU time and 30~GB of memory
    can solve linear-chain workflows of up to 14 components across eight sites.
\end{abstract}

\begin{links}
    \link{Domain}{https://github.com/taylorpaul/WORKSWORLD}
    \link{Benchmarks}{https://gitlab.com/thpaul/worksworld-benchmarks/}
    \link{Data}{https://gitlab.com/thpaul/worksworld-benchmarks/-/jobs/13325198684/artifacts/download}
\end{links}

\section{Introduction}

Spending billions of dollars on data platforms and \ac{ai}, most organizations
struggle to see any tangible \ac{roi}~\cite{challapally2025genai}. Why? Today's
platforms and algorithms deliver little value without unfettered access to
collected, curated, and current data~\cite{Heck2024What} from the distributed
systems that support daily operations. Establishing effective data pipelines, or
\emph{data-intensive workflows} (hereafter, \emph{workflows}), persists as a
prerequisite to productive employment of \ac{ai} tools. Such workflows must
process diverse data sources into required formats and feed them into multiple
systems and models without excessive cost or prohibitive latency. Optimally
planning and scheduling such workflows, especially those sufficiently complex to
meet real-world use cases, most often proves computationally
intractable~\cite{Versluis2021survey, Ghallab1998PDDL}.

\begin{figure}[t]
    \centering
    \resizebox{0.75\linewidth}{!}{
        \begin{tikzpicture}[
                node distance=1.5cm,
                box/.style={rectangle, draw, thick, minimum width=2.5cm, minimum height=1cm, align=center},
                smallbox/.style={rectangle, draw, thick, minimum width=1.8cm, minimum height=1cm, align=center},
                arrow/.style={->, thick, >=stealth'}
            ]

            \node[box, fill=tikzYellow] (converter) {YAML to PDDL\\ Converter};

            \node[box, fill=tikzYellow, right of=converter, node distance=4cm] (generator) {YAML Generator\\{\footnotesize (environment \& goal)}};

            \node[smallbox, below of=converter, xshift=-1cm, fill=tikzGreenLight] (problem) {Problem\\ PDDL};
            \node[smallbox, right of=problem, node distance=2cm, fill=tikzGreenLight] (domain) {Domain\\ PDDL};

            \node[box, below of=problem, xshift=1cm] (planner) {Existing Numeric\\ Planner(s)};

            \node[box, below of=planner] (plan) {Validate\\Output Plan};

            \node[box, right of=plan, node distance=4cm, fill=tikzYellow] (viz) {Visualization\\{\footnotesize (initial state + result)}};

            \node[box, right of=domain, node distance=3cm] (yaml) {YAML Config\\{\footnotesize (environment \& goal)}};

            \node[coordinate] (user) at ([yshift=-1.1cm]yaml) {};
            \draw[thick] (user) circle (0.111cm);
            \draw[thick] ([yshift=-0.111cm]user) -- ++(0, -0.311);
            \draw[thick] ([yshift=-0.155cm]user) -- ++(-0.178, -0.133);
            \draw[thick] ([yshift=-0.155cm]user) -- ++(0.178, -0.133);
            \draw[thick] ([yshift=-0.4cm]user) -- ++(-0.133, -0.222);
            \draw[thick] ([yshift=-0.4cm]user) -- ++(0.133, -0.222);
            \node[below] (engineer-label) at ([yshift=-0.622cm]user) {\footnotesize Data Engineer};

            \begin{pgfonlayer}{background}
                \node[draw, dashed, thick, tikzGreenDark, rounded corners=4pt,
                    inner sep=0.2cm, fill=tikzGreenFaint,
                    fit=(problem)(domain)] (formulation-box) {};
            \end{pgfonlayer}
            \node[rotate=90, font=\footnotesize, anchor=center, align=center]
            at ([xshift=-0.4cm]formulation-box.west) {General\\Formulation};

            \draw[arrow] ([yshift=0.3cm]user) -- (yaml);
            \draw[arrow] (yaml) -- (converter);
            \draw[arrow] ([xshift=-0.8cm]converter.south) -- ([xshift=-0.8cm]converter.south |- problem.north);
            \draw[arrow] (problem) -- (planner);
            \draw[arrow] (domain) -- (planner);
            \draw[arrow] (planner) -- (plan);
            \draw[arrow] (plan) -- (viz);
            \draw[arrow] (generator) -- (converter);
            \draw[arrow] (viz) -- (engineer-label.south);

        \end{tikzpicture}
    }\caption{The \worksworld{} framework. Primary research contributions are
        green; secondary engineering artifacts provided without experimental
        evaluation are yellow. }\label{fig:framework}
\end{figure}
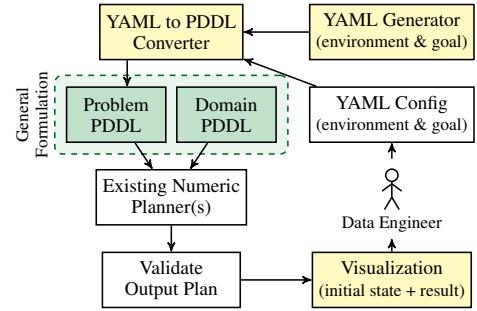

Diverse communities have long pursued efficient placement of computational tasks
on distributed heterogeneous resources: \ac{hpc}~\cite{da2024workflows},
distributed computing~\cite{Dustdar2023Distributed}, and machine
learning~\cite{Kreuzberger2023Machine}. These communities share a core problem,
yet bring distinct perspectives, objectives, and assumptions that shape both
tractability and real-world applicability of solutions. Prior work typically
begins with a fully-formed \emph{workflow}---a \ac{dag} of computational tasks
connected by data dependency edges---and solves a scheduling problem to place
tasks on compute resources while optimizing one or two metrics (\eg, latency,
throughput, or cost)~\cite{Vivas2024Trends}. We jointly plan and schedule a
\ac{dag} of both processing and data components, explicitly decoupling compute,
storage, and network decisions.

Data-intensive workflows have received increased attention over the past
decade~\cite{Suter2023Driving, Heck2024What}, inspiring new organizational
approaches (\eg{} data mesh~\cite{goedegebuure2024data}) and roles (\eg{} data
engineers). Complexity in planning these workflows increases when data producers
(\emph{sources}) and consumers (\emph{sinks}) span multiple networks or
\emph{sites}, especially when consumers require the same source data in
different formats. A data engineer must decide which data conversions to apply,
whether to process data centrally (\ie{} the cloud) or simultaneously at
multiple sites, and how to place processing tasks across available compute
resources while meeting capacity and interoperability constraints. This
complexity, combined with the workflow graph's direct impact on performance and
cost, makes such workflows strong candidates for joint planning and scheduling
approaches~\cite{long2023scheduling}.

To aid data engineers in their work, we develop the framework depicted in
Figure~\ref{fig:framework} that permits them to define the problem environment,
workflow components, and goals (output data formats and locations) as a
\ac{yaml} configuration file. Our framework converts the \ac{yaml} to a
\ac{pddl} problem and solves a numeric planning problem using existing numeric
planners and a new \ac{pddl} domain we call \worksworld{}. We validate the
output plan\footnote{https://github.com/KCL-Planning/VAL}~\cite{howey2004val}
and provide a simple visualization tool depicting both the provided initial
state and result of executing the plan. In contrast, existing solutions are
largely proprietary and platform-specific~\cite{amado2025automated,
    sohrabi2013htn}. To our knowledge, no publicly available system addresses our
exact problem formulation. Thus, we demonstrate viability through scalability
results rather than a competitive baseline comparison. We find that a \ac{soa}
numeric planner, the \ac{enhsp}, can plan and schedule satisficing workflows of
seven processing and seven data components across eight sites with reasonable
CPU time (one hour) and memory (30~GB) on commodity hardware.


\section{Problem and Motivation}\label{sec:problem}

\begin{figure}[t]
    \centering
    \includegraphics[width=0.9\linewidth]{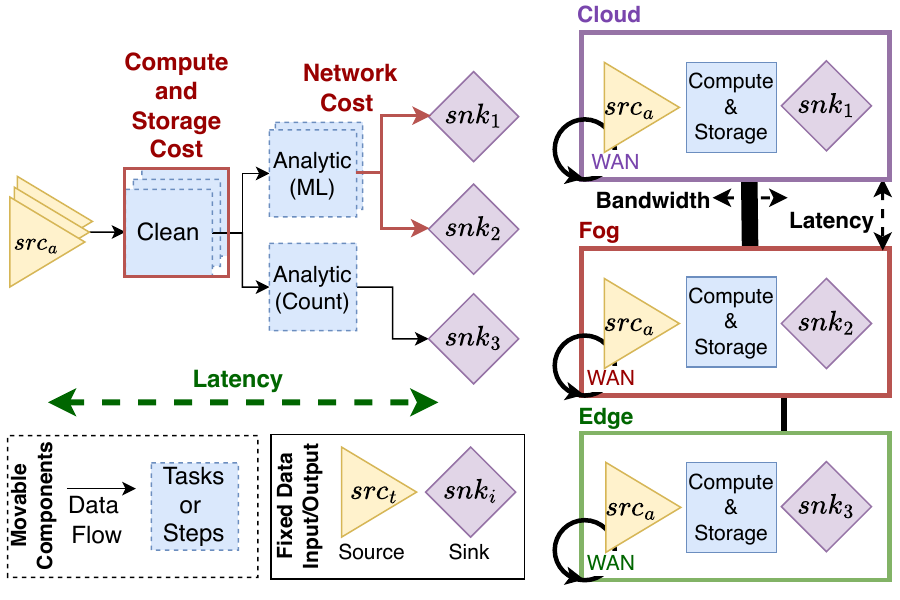}
    \caption{Our scheduling problem consists of mapping movable components in the left-hand \ac{dag}
        to utilize compute, storage and network resources across sites.}\label{fig:problem}
\end{figure}

Depicted in Figure~\ref{fig:problem}, we seek to map pipelined workflows (see
Definition~\ref{def:pipelined}) \emph{permanently}, until deprovisioned, to
compute and storage resources across distributed cloud, fog, and edge sites~(see
Section~\ref{subsec:dist_comp_env}). We do not require the entire workflow graph
as input and thus solve an integrated planning and scheduling problem. We seek
satisficing mappings of workflows to site resources that minimize compute,
storage and network costs while constraining latency.

\subsection{Motivating Workflow Examples}

We present three workflows spanning distinct cost-latency tradeoff regimes
to illustrate the value of decoupling processing, data placement, and routing.

\begin{enumerate}
    \item \emph{Archival Workflow:} Digital media archiving~\cite{hou2006scientific}.
          Records from three distributed sites must be encoded, compressed, and
          stored at a central cloud site (hours tolerable). Tradeoff:
          compress at source (less transfer cost) or centrally to consolidate
          compute?
    \item \emph{Sensor Workflow:} Wildfire detection~\cite{Altintas2022Towards}.
          Edge sensors generate raw data requiring pre-processing before
          algorithm inference to detect smoke and trigger simulations (minutes
          tolerable before human review). Tradeoff: transfer raw data to
          the cloud, or move pre-processing and inference to edge sites?
    \item \emph{Automation Workflow:} Cybersecurity. Firewall events undergo
          normalization and anomaly detection to trigger automatic IP blocking
          (sub-second latency required). Tradeoff: meeting strict latency bounds
          while placing compute and data to minimize cost.
\end{enumerate}

These workflows span the cost-latency tradeoff space relevant to our research:
workflow purpose drives latency tolerance, while data volume and compute
placement decisions drive cost. They motivate our research and guide our
experimental setup in Section~\ref{sec:methodology}.

\section{Background}\label{sec:background}

Now we lay down a foundation, built by others, on which our contribution rests.
We briefly define the computing environment in which we plan, scope our workflow
scheduling problem, and provide the required \ac{ai} planning prerequisites.

\subsection{Compute Environment}\label{subsec:dist_comp_env}

The distributed computing research community provide abstractions to describe
the computing environment they study. The early 2000's saw a heavy focus on Grid
Computing~\cite{Foster2001Anatomy} before cloud-computing~\cite{Mell2011NIST}
stole much of the community's focus. When latency sensitive workflows did not
fit the cloud only approach, the Cloud-Compute
Continuum~\cite{Moreschini2022Cloud} emerged to define three general categories
of compute resources: cloud, fog and edge which we leverage in
describing different sites and their compute, storage and network resources.

\subsection{Pipelined Workflows Scheduling}

Scheduling compute tasks across available resources proves an incredibly broad
topic studied by multiple communities across many
decades~\cite[Ch.~15]{Ghallab2004Automated}. There is no best scheduling
approach in general to workflows due to the NP-hardness of the problem and the
influence of specific scheduling context on the time and space required for
finding acceptable solutions~\cite{Versluis2021survey}. We find a scoping to
pipelined workflows scheduling clarifying in terms of our context; see
Definition~\ref{def:pipelined}.

\begin{definition}[Pipelined Workflow Scheduling]\label{def:pipelined}
    Pipelined workflow scheduling encompasses executing the same workflow
    on varying datasets of identical size across parallel machines utilizing task,
    data, pipelined and/or replicated parallelism.~\cite{benoit2013survey}
\end{definition}

In our setting, the datasets are individual or groups of source data messages
aggregated at our input data components. Benoit
\etal{}~\citeyear{benoit2013survey} provide a survey and taxonomy to classify
pipelined workflow scheduling characteristics and specify complexity for
scheduling linear task graphs. With their taxonomy, we classify our problem as
scheduling pipelined workflows across related-heterogenous compute and storage
resources at distributed sites connected by structured-heterogenous, bandwidth
bounded, network links. We employ a multi-objective performance model that
minimizes resource utilization (\ie{} cost) while constraining latency
(corresponding to \emph{makespan} in single dataset \ac{dag} scheduling).
However, our \emph{permanent} scheduling of components to resources rather than
repeatedly scheduling dataset execution tasks removes the temporal sequencing that
drives the scheduling complexity in classical pipelined workflows. The hardness
of \worksworld{} instead arises from multi-resource placement under network
constraints, which proves structurally similar to the NP-hard virtual network
embedding problem~\cite{fischer2013virtual}.

\subsection{Classical and Numeric
    Planning}\label{subsec:planning}

\ac{ai} Planning, or automated planning and scheduling, consists of leveraging
computers and algorithms to generate plans to execute towards achieving a set of
goals; Ghallab, Nau and Traverso~\shortcite{Ghallab2004Automated} cover in
detail topics we discuss briefly.

\emph{Classical planning} possesses the most restrictive assumptions but
maintains a general complexity of
PSPACE-complete~\cite{bylander1994computational}; it does not permit reasoning
about numeric variables. Planning systems that interact with real-world physical
systems often require \emph{numeric planning}~\cite{scala2016interval}. Numbers
increase the complexity in the worst case to semi-decidable and removes bounds
on plan length possible in propositional planning~\cite{haslum2019introduction}.
Even so, we leverage numeric planning in our work due to our foundational
dependency on numeric measurements of compute, storage and network resource
availability. Due to the \emph{permanent} nature of our data pipelines, we do
not leverage \emph{temporal planning} to reason about time as the ordering of
actions (\ie{} sequential planning) to schedule and connect components suffices.

In the past three decades, the \ac{ai} planning community produced several
solvers capable of planning with numerics: MetricFF~\cite{hoffmann2003metric},
LP-td~\cite{gerevini2006approach}, SMTPLAN\(^+\)~\cite{cashmore2016compilation}.
We exclusively evaluate
\ac{enhsp}\footnote{https://sites.google.com/view/enhsp/} due to the
functionality it supports (\ie{} Quantifiers in pre-conditions and goals), its
support for parsing large problem files and domains (resp. MetricFF), and its
effective pruning of the search space with \ac{aibr}
preprocessing~\cite{scala2016interval}. We leave finding additional compatible
planners and testing them for future work (\eg{} SMTPLAN\(^+\),
GOOSE~\cite{chen2024graph}).

With our scope established, we turn to defining the general numeric planning
problem we solve. We reproduce others' work~\cite{Fox2003PDDL2,
    scala2016interval, scala2020hmrp} minimally for the readers' benefit. The state
of our system (\(s_i \in \mathcal{S}\)) amounts to a total assignment of a set
of propositional variables (\(\mathcal{F}\)) and a partial assignment of numeric
variables (\( \mathcal{X} := \{x \mid x \in \mathbb{R} \} \)); let the set of
all variables \(\mathcal{V} = \mathcal{F} \cup \mathcal{X} \). Propositional
conditions are positive literals (\(p = \top \)) or negative literals (\(p =
\bot \)) where \(p \in \mathcal{F}\). Numeric conditions take the form \(\langle
\xi,\, \trianglerighteq,\, l \rangle \) where \( \xi \) is a numeric expression
over a subset of \( \mathcal{X} \), \(\trianglerighteq \in \{ \leq, <, =, >,
\geq \} \) and \(l\in\mathbb{Q}\). The operators permitted in these expressions
are specific to the planner (or heuristic) leveraged, we currently leverage the
standard set of operators \( +, -, \times, \div \) in \worksworld; \ac{enhsp}
supports more. With the basics covered,
Definition~\ref{def:numeric_action} describes a numeric action and
Definition~\ref{def:numeric_prob} formulates the numeric planning problem.

\begin{definition}[Numeric Action]\label{def:numeric_action} A numeric action
    \(a\) consists of two sets: \(\text{pre}(a)\) the set of numeric and
    propositional pre-conditions permitting a planner to select an action and
    \(\text{eff}(a)\) the set of propositional and numeric effects that define
    the resulting state changes after executing the action (\( a =
    \langle \text{pre}(a),\, \text{eff}(a) \rangle \)). Propositional effects set
    (\( p = \top \))  or unset (\( p = \bot \)) a proposition \(p \in
    \mathcal{F}\). Numeric effects can assign (\(=\)), increase
    (\(\mathrel{+}=\)) or decrease (\(\mathrel{-}=\)) numeric variables \(x \in
    \mathcal{X}\) by \( \xi \).  An action's \(\text{eff}(a)\) must only update a
    variable \(x\) \emph{at most} once.
\end{definition}

\begin{definition}[Numeric Planning Problem]\label{def:numeric_prob} A numeric
    planning problem (\( \, \Pi = \langle s_0, \mathcal{A}, \mathcal{G},
    \mathcal{V} \rangle \)) consists of an initial state
    (\(\boldsymbol{s_{0}}\)), which is an initial assignment (complete for
    \(\mathcal{F}\)) to the variables in \(\mathcal{V} = \mathcal{F} \cup
    \mathcal{X} \),  a set of numeric actions (\(\mathcal{A}\)), and a set of
    propositional or numeric goal conditions (\(\mathcal{G}\)).
\end{definition}

The act of planning involves selecting a sequence of actions that transitions
our system from the initial state \(s_0\) to a goal state \(s_g\) that satisfies
(\( \, \models \, \)) all \( g \in \mathcal{G}\) or \(s_g \models \mathcal{G}\).
This sequence of actions is the planner's output plan \( \pi \). We reproduce
Definition~\ref{def:plan} from~\cite{scala2016interval} with a slight
modification to the description of an optimal plan; see their work for more
details on the applicability of numeric actions in a state.

\begin{definition}[Plan]\label{def:plan} A plan \( \pi \) for \(\Pi = \langle s_0,
    \mathcal{A}, \mathcal{G}, \mathcal{V} \rangle \) is a sequence of actions
    \(a_0, \ldots, a_{n-1}\) from \(\mathcal{A}\) such that each action in \( \pi \)
    is applicable in the state resulting from the application of its
    predecessors, i.e., \(s_0 \models \text{pre}(a_0)\), \(\text{succ}(s_0,
    a_0) \models \text{pre}(a_1)\), etc, and \(\text{succ}(s_{n-1}, a_{n-1})
    \models \mathcal{G}\). A plan \( \pi \) \textit{is said to be optimal if,
        among all valid plans, it has a minimal number of actions (by default)
        or it minimizes or maximizes some specified \(x \in \mathcal{X}\) or
        \(\xi \) over \( \{x\} \).}
\end{definition}

With our problem and desired solution sufficiently specified, we turn to
describing the required inputs for domain-independent planners.

\subsection{Planning Inputs}\label{subsec:pddl}

The ability to encode a planning problem into a common format permits the
creation of domain-independent solvers; a single solver for many types, or
domains, of problems.
The \ac{ai} Planning Community produced several versions of
\acrfull{pddl} to standardize inputs across
planners~\cite{Ghallab1998PDDL,Fox2003PDDL2,Edelkamp2004PDDL22,Gerevini2005Plan}
We encode the \worksworld{} in \ac{pddl}~2.1 level 2~\cite{Fox2003PDDL2} which
permits our use of numeric fluents (or functions); we do not leverage the
durative actions of level 3. For an in-depth overview of \ac{pddl}
see~\cite{haslum2019introduction}.

To solve the planning problem of Definition~\ref{def:numeric_prob}, \ac{ai}
planners require a planning instance (\(\mathcal{I} = \langle \text{Domain},
\text{Problem} \rangle \)) encoded in \ac{pddl} as input. While the domain gets
defined once, generating problem instances in \ac{pddl}, especially when \(s_0\)
or \(\mathcal{V}\) prove large, remains a well-known barrier to users leveraging
\ac{ai} planners. Therefore we permit users (or systems) to more concisely
define the problem instance using a popular human-readable data serialization
language (\acrshort{yaml}\footnote{https://yaml.org/}) to produce a problem
configuration file with common units and minimal requirements. We leverage a
templating library
(Jinja\footnote{https://jinja.palletsprojects.com/en/stable/}) and python
scripts to produce problem instances in PDDL from the supplied configuration
file. The \ac{pddl} contains additional values from calculations across input
values as well as consistent units to minimize mathematical operations required
during planning (\ie{} storage in MB and bandwidth in MB/s); see example YAML
and \ac{pddl} files available
online\footnote{https://github.com/taylorpaul/WORKSWORLD}.

\section{General Problem Formulation}\label{sec:general_problem}

\begin{figure}[t]
    \centering
    \includegraphics[width=0.75\linewidth]{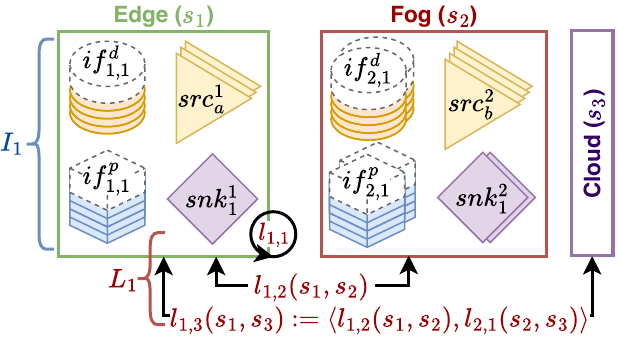}
    \caption{Sites and the relevant annotations and resources for our scheduling problem.}\label{fig:sites}
\end{figure}

We start by defining our environment, which informs our resource graph, and the
workflows we seek to efficiently distribute. Figure~\ref{fig:sites} depicts our
sites and their main resources. The \emph{site}, see Definition~\ref{def:site},
proves the foundational abstraction of our environment. We assume that an
organization manages specific sites and peers them, bringing them from the set
of globally available sites (\(\mathbb{S}\)) into a subset (\(S^{peers}_m\))
considered for scheduling; see Equation~\ref{eqn:peers}. Peered sites must have
at least one link connecting to the set of peers. These logical links (\eg{} in
a Software-Defined WAN) can be direct, or composite consisting of two hops.
These sites possess data \emph{sources} of a specific data type
(\(\text{src}_t\)) that serve as inputs to our workflows. They also may have
\emph{sinks} which are applications that consume workflow outputs.

\begin{equation}\label{eqn:peers}
    S^{peers}_m  := \{s_n \,|\, s_n \in \mathbb{S} \text{ and } \exists \, l(s_m, s_n) \}           \\
\end{equation}

\begin{definition}[Site]\label{def:site} We define a \emph{site} (\(s_m\)) that
    exists in the set of peered sites (\(S^{peers}_m \)) as a list of three
    sets: the set of interfaces (\(I_m\)) located at the site, a set of
    relations (\(L_m\)) to describe intrasite and intersite network
    connections to peered sites, and the set of site-level annotations
    (\(A_{m}\)) to capture properties of the site where (indices \(m,i,k
    \in\mathbb{N}\)):
    \begin{equation}
        \begin{aligned}
            s_m & := \left \{ \, I_m, L_m, A_m\,\right \} \text {such that }                              \\
                & I_m  := \left \{ \, if_{m,i} \, \right \}                                               \\
                & L_m  := \left \{ \, l_{m,k}(s_m, s_n) \,\middle|\, s_m, s_n \in S^{peers}_m \,\right \} \\
                & A_m  := \left \{ \text{ properties of site \emph{m} } \right \}
        \end{aligned}
    \end{equation}
\end{definition}

For this work, we present an interface abstraction to expose compute and storage
resources for scheduling at sites. Thus, we schedule on a single interface over
a clustered set of resources and simplify the number of metrics we must track.
We identify two types of interfaces: \emph{data sharing} (\(d\)) and \emph{data
    processing} (\(p\)). We envision \emph{\acp{dsi}} to have different performance
characteristics and resource costs (\ie{} disk, CPU, RAM). There are existing
open-source tools that an organization could standardize across sites for
streaming, object storage or file sharing (\eg{} Apache Kafka, CEPH, SeaweedFS,
\etc).\ \Acrfullpl{dpi} prove most similar to resource managers like Kubernetes,
Docker Swarm, or Slurm. Making these interfaces available as network-level
resources for applications to leverage requires changes to the current network
model (TCP/IP) which many others explore~\cite{Zhang2014Named,
    Krol2019Compute,paul2024accelerating}.

\begin{definition}[Interface]\label{def:interface} We define an \emph{interface}
    (\(if\)) as a network-level resource to schedule workflow components with a
    type (\(t\)): data sharing (\(d\)) or processing (\(p\)). A set of metrics
    (\(IF^{R}\)) describes the total and available resources. (\(IF^{C}\)) is
    the set of components instantiated on the \(if\) and a set
    of annotations (\(IF^{A}\)) describes hardware, software and performance
    characteristics. We define (\(m,i, \in \mathbb{N}\)):
    \begin{equation}
        \begin{aligned}
            if^{t}_{m,i} & := \left \{ IF^{R}_{m,i},\, IF^{C}_{m,i},\, IF^{A}_{m,i} \right \}
            \text{ such that } t \in \{d,p\}                                                  \\
        \end{aligned}
    \end{equation}
\end{definition}

We establish our workflow components in Definition~\ref{def:wf_component} that
we will schedule on our interfaces and our rules for connecting them into a
workflow; see Definition~\ref{def:workflow}.

\begin{definition}[Workflow Component]\label{def:wf_component} We define a
    \emph{workflow component} (\(wfc^{c,t}\)) as a logical boundary of work
    required by a workflow, delineating two classes of work: data processing
    (\(p\)) or data sharing (\(d\)) of a specific type (\(t\)) from the set of
    known component types (\(\mathbb{T}_c\)), such that (\(x,y\in\mathbb{N}\)):
    \begin{equation}
        \begin{aligned}
             & wfc_{x}^{c,t}  := \{ \, C^{R}_{c,t},\; C^{IO}_x,\; C^{A}_{t,x} \mid c\in \{p,d\}, t\in\mathbb{T}_c\}                             \\
             & C^{R}_{c,t}    := \{ \, amt^{r}_{c,t} \mid r \in \{\text{comp, store, mem, bw}\} \}                                              \\
             & C^{IO}_{x}     := \{C^{in}_{y \xrightarrow[l]{} x}, C^{out}_{x \xrightarrow[l]{} y} \mid y\neq x,\, l \in L_m \text{ at } s_m \} \\
             & C^{A}_{t,x}    := \{ \,\text{properties of } t \text{ or } x\, \}
        \end{aligned}
    \end{equation}
\end{definition}


\begin{definition}[Workflow]\label{def:workflow} We define a \emph{workflow} as
    a \ac{dag} of \emph{workflow components} with components \emph{only}
    connecting via a \emph{link} to one or more components of another class; \(d \to p\)
    or \( p \to d \) but \emph{never} \( d \to d\). A workflow
    \emph{must} start and end with one or more data components.
\end{definition}

With these definitions in hand, we prove prepared to pursue a planning
and scheduling approach that can implement our model and produce actions
required to map our workflows onto our connected site resources.

\section{Methodology}\label{sec:methodology}

We document the \worksworld{} domain and describe its complexity. Then we
describe our research questions and corresponding experiments.

\subsection{\worksworld{}}\label{sub:worksworld}

Our main contribution is a new numeric planning domain, \worksworld{}, that
permits automated planning of pipelined workflow graphs to process and deliver
data in the correct format to the desired site. We describe the domain's primary
objects and actions leveraging a single-site representation and discuss the
added complexity when growing the resource and workflow graph. We developed and
tested several iterations of the domain to achieve what we believe represents
the right balance of expressivity, realism and complexity.

\subsubsection{Object Hierarcy}

\begin{figure}[t]
    \centering
    \resizebox{0.95\columnwidth}{!}{%
        \begin{tikzpicture}[
                font=\sffamily\scriptsize,
                basic/.style={
                        draw=black!60,
                        thick,
                        rounded corners=2pt,
                        align=center,
                        inner sep=2pt,
                        minimum height=0.4cm,  
                        drop shadow={opacity=0.2, shadow xshift=1pt, shadow yshift=-1pt}
                    },
                root/.style={basic, fill=white, font=\sffamily\bfseries},
                instance/.style={basic, fill=oiSkyBlue!40},
                descriptor/.style={basic, fill=oiOrange!40},
                site/.style={basic, fill=oiGray!50},
                edge from parent/.style={
                        draw=black!50,
                        thick,
                        latex- 
                    },
                grow=down,
                level 1/.style={sibling distance=1.5cm, level distance=0.75cm},
                level 2/.style={sibling distance=1.6cm, level distance=0.75cm},
                level 3/.style={sibling distance=0.8cm, level distance=0.75cm}
            ]

            \node[root] {Object}
            child { node[instance] {Instance}
                    child { node[instance] {Link}
                            child { node[instance] {DL} }
                            child { node[instance] {CL} }
                        }
                    child { node[instance] {Interface}
                            child { node[instance] {DSI} }
                            child { node[instance] {DPI} }
                        }
                    child  [sibling distance=1.6cm]{ node[instance] {WC}
                            child { node[instance] {DC} }
                            child { node[instance] {PC} }
                        }
                }
            child [sibling distance=1.8cm] { node[site] {Site}  }
            child [sibling distance=1.2cm] { node[site] {Resource}  }
            child { node[descriptor] {Descriptor}
                    child [sibling distance=1.5cm]{ node[descriptor] {WCT}
                            child { node[descriptor] {DCT} }
                            child { node[descriptor] {PCT} }
                        }
                    child [sibling distance=1.cm] { node[descriptor] {DIR} }
                };
        \end{tikzpicture}
    }
    \\
    \vspace{0.5em}
    \scriptsize
    \begin{tabular}{@{}l@{\hspace{3em}}l@{\hspace{3em}}l@{}}
        I: Instance           & L: Link            & DL: Direct Link      \\
        CL: Composite Link    & IF: Interface      & DSI: Data Sharing IF \\
        DPI: Data Process. IF & WF: Workflow       & WC: WF Component     \\
        DC: Data Comp.        & PC: Process. Comp. & S: Site              \\
        R: Resource           & WCT: WC Type       & DCT: DC Type         \\
        PCT: PC Type          & DIR: Direction     &                      \\
    \end{tabular}
    \caption{PDDL type hierarchy for \worksworld{} and abbreviations utilized throughout paper.}\label{fig:domain-types}
\end{figure}
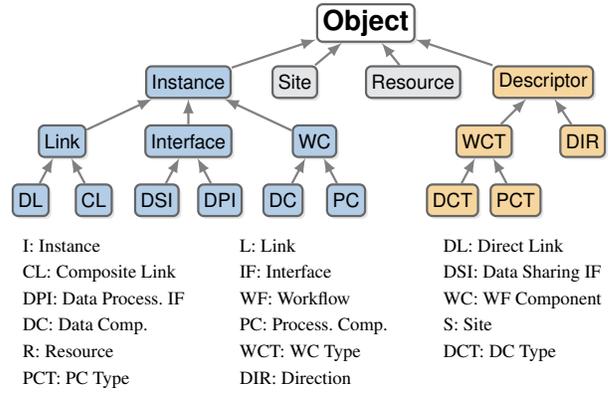

We introduce 17 object types to implement our model from
Section~\ref{sec:general_problem} in our \ac{pddl} domain; see
Figure~\ref{fig:domain-types}. For our resource graph we define types
\emph{sites}, connected via \emph{links} (direct and composite),
\emph{interfaces} (\ac{dsi}/\ac{dpi}) and \emph{resource}. We leverage constants
of type resource in our domain: \emph{storage}, \emph{compute}, \emph{network}
and \emph{config}. These constants describe resources available on (links and
interfaces) or required (workflow components) by \emph{instance} subtypes. We
employ the config resource to reason about whether or not we must copy a
workflow component configuration to a site or if it already exists there. Thus we can
account for the cost of copying configurations like container images or model
weights that may prove prohibitively large. We also provide the domain constants
\emph{input} and \emph{output} of type \emph{direction}, always relative to a
processing component. Figure~\ref{fig:domain-types} also contains the
abbreviations for our object types we employ throughout the paper.


\subsubsection{Predicates (\(\mathcal{F}\))}

Table~\ref{tab:predicates-fluents} lists our 12 predicates; we discuss the most relevant
here and point the reader to comments in the \ac{pddl} domain for details. The
\texttt{available\_at} predicate declares what links or interfaces exist at a
site. We also use the predicate to describe if a configuration for a workflow
component is available at a site. The \texttt{fixed} predicate marks workflow
components that the planner cannot schedule. We use this for source and sink
data components fixed to specific sites, while \texttt{\(\neg\)fixed} components
can be scheduled at multiple sites. The \texttt{has\_data} predicate tracks data
provenance through a planned workflow graph. In our goals, it describes what
source data component we want processed into the format required by the
destination sink data component.


\begin{table}[t]
    \centering
    {\scriptsize 
        \begin{tabular}{@{}l@{\hspace{0.5em}}l@{}}
            \textbf{Predicates (\(\mathcal{F}\))}    & \textbf{Numeric Fluents (\(\mathcal{X}\))} \\
            \midrule
            \texttt{(available\_at I R S)}           & \texttt{(resource\_total I R)}             \\
            \texttt{(scheduled\_on WC IF)}           & \texttt{(resource\_available I R)}         \\
            \texttt{(type\_of WC WCT)}               & \texttt{(work\_amount WC R)}               \\
            \texttt{(fixed WC)}                      & \texttt{(msg\_max\_rate WC)}               \\
            \texttt{(linked L S S)}                  & \texttt{(msg\_actual\_rate WC IF)}         \\
            \texttt{(link\_uses CL DL)}              & \texttt{(msg\_size DCT)}                   \\
            \texttt{(connected L PC DPI DC DSI DIR)} & \texttt{(network\_latency L)}              \\
            \texttt{(has\_input WC IF)}              & \texttt{(work-cost-weight R)}              \\
            \texttt{(input\_format PC DCT)}          & \texttt{(total-cost)}                      \\
            \texttt{(output\_format PC DCT)}         & \texttt{(absolute-latency)}                \\
            \texttt{(processed\_by DC DSI PC DPI)}   &                                            \\
            \texttt{(has\_data DC DSI DC DSI)}       &                                            \\
        \end{tabular}}
    \caption{Available predicates and numeric fluents in \worksworld{}.}\label{tab:predicates-fluents}
\end{table}

\subsubsection{Numeric Fluents (\(\mathcal{X}\))}

We leverage ten numeric fluents in \worksworld{} depicted in
Table~\ref{tab:predicates-fluents}. Links and interfaces must have
\texttt{resource\_total} and \texttt{resource\_available} fluents assigned in the
initial state; \texttt{work\_amount} fluents describe how many resources
(compute cores or storage bytes) scheduling a component on an interface
consumes. We multiply \texttt{msg\_max\_rate} times \texttt{msg\_size} to
consume permanent bandwidth resources when connecting components across links.
We will discuss the other relevant numeric fluents when we discuss
\worksworld{}'s cost and performance calculations.

\subsubsection{Actions (\(\mathcal{A}\))}

Our action definitions prove the most influential on the compute and memory
requirements during planning. The number of objects input into our actions
greatly affects the grounding time; the number of actions affects \ac{enhsp}
\ac{aibr} preprocessing step as it loops over each
action~\cite{scala2016interval} to prune the state space. As we show later,
these times dominate our planning time as problem sizes grow due to the size of
the state space. Table~\ref{tab:actions} presents our six actions and the
objects they require as inputs. We briefly walkthrough how a planner leverages
our actions during planning before discussing how the actions affect cost and
performance calculations.

Starting from \(s_0\), the planner builds a workflow graph satisfying
\(\mathcal{G}\) by iteratively scheduling and connecting components. To
\texttt{schedule\_component}, the component's configuration must exist at the
target site; \texttt{replicate\_code} copies configurations from peered sites as
needed. After scheduling required data and processing components, the planner
connects them via \texttt{connect\_direct\_link} or
\texttt{connect\_composite\_link} based on available links, bandwidth, and
latency. It then selects \texttt{propagate\_input} or \texttt{propagate\_output}
depending on the connection \emph{direction} relative to the processing
component, recording that a downstream component \texttt{has\_data} from an upstream
source. This process repeats until the planner achieves all goal conditions or
the planner exhausts time limits, memory, or the entire pruned state space.

Separating the propagate effects from our connect actions provides two benefits.
First, we avoid conditional effects for input or output in the connect actions,
which the specific \ac{enhsp} heuristics (\ie{} \ac{mrp}~\cite{scala2020hmrp})
we require do not
support\footnote{https://sites.google.com/view/enhsp/home/how-to-use-it}.
Second, we avoid requiring input and output versions of the more complex connect
actions which reduces the required object combinations in actions during
grounding; see object inputs in Table~\ref{tab:actions}.

\begin{table}[t]
    \centering
    {\footnotesize
        \begin{tabular}{l}
            \textbf{Actions (\(\mathcal{A}\)) \& Input Object Types}  \\
            \midrule
            \texttt{(replicate\_code WC S S L)}                       \\
            \texttt{(schedule\_component WC R IF R S)}                \\
            \texttt{(connect\_direct\_link}                           \\
            \texttt{\hspace{2em} PC DPI S DC DCT DSI S DL DIR)}       \\
            \texttt{(connect\_composite\_link}                        \\
            \texttt{\hspace{2em} PC DPI S DC DCT DSI S CL DL DL DIR)} \\
            \texttt{(propagate\_input PC DPI DC DSI DC DSI)}          \\
            \texttt{(propagate\_output PC DPI DC DSI DC DSI)}         \\
        \end{tabular}}
    \caption{Available actions in \worksworld{} to build and schedule pipelined workflows.}\label{tab:actions}
\end{table}

\subsubsection{Cost and Performance}\label{subsub:cost}

In \worksworld{} we minimize on cost (see Definition~\ref{def:total-cost}) and
constrain on a message latency metric (see Definition~\ref{def:absolute}) in
\(\mathcal{G}\). Our use of a domain-independent planner permits adapting these
metrics with ease compared to domain specific optimization algorithms. We
increase the cost metric for every action in Table~\ref{tab:actions} except the
propagate actions. Importantly, we avoid utilizing \texttt{resource\_available}
to avoid state-dependent action costs; which \ac{enhsp} only
experimentally supports. We instead ensure links and interfaces do not get over
provisioned via \(\text{pre}(a)\) checking for sufficient
\texttt{resource\_available}.

\begin{definition}[Total Cost]\label{def:total-cost} We define
    \emph{total-cost} as a weighted (\emph{work-weight R}) resource
    utilization (\emph{work\_amount WC R}) normalized by resource total
    (\emph{resource\_total I R}) on a link or interface.
\end{definition}

For our latency measure, due to an inability to track the bottleneck path and
report the worst case latency, we instead define \texttt{absolute\_latency} in
Definition~\ref{def:absolute}. We increase this numeric fluent in every action
that establishes a permanent part of the workflow: the schedule and connect
actions. Again, we apply a similar strategy to avoid
state-dependent calculations by calculating latencies with a constant numeric
fluent (\texttt{msg\_max\_rate}). We avoid more detailed definitions of cost and
latency here due to space but point a curious reader to the
domain \ac{pddl} for details.


\begin{definition}[Absolute Latency]\label{def:absolute}
    We define \emph{absolute-latency} as the sum of the time (seconds) it takes a single message
    to traverse every edge of the workflow graph.
\end{definition}

\subsubsection{\worksworld{} Expressivity}

\worksworld{} generalizes to any workflow expressible as a \ac{dag} of
alternating data sharing and processing steps where components remain online
until deprovisioned. This covers data pipelines common in sensing, scientific
computing, and machine learning. It does \emph{not} cover short-lived batch jobs
that maximize compute utilization or minimize makespan across temporal tasks
(\eg{} some HPC workloads).

\subsubsection{State Space Complexity}\label{subsub:Complexity}

We present Table~\ref{tab:object-complexity} on the grounded state-space
complexity increase as our problem sizes grow. We declare objects
types that have multiple inputs to the same action (see Table~\ref{tab:actions})
have the potential to impact the state space most significantly. While this
table helps understand object-level complexity, there are combinatorial effects
when we add sites and/or workflow components. Table~\ref{tab:grounding} presents
the state-space increase when adding sites, with their interfaces and links, and
also when adding an extra data and processing component. Fortunately, \ac{enhsp}
smartly prunes this state space before initializing search to permit
sufficiently large problem sizes. We explore this in our experiments next.

\begin{table}[t]
    \centering
    {\footnotesize
        \begin{tabular}{lccl}
            \textbf{Object Type}          & \textbf{F} & \textbf{X} & \textbf{A}           \\
            \midrule
            Site                          & 0          & 0          & \(\mathcal{O}(n^2)\) \\
            Direct Link                   & 2          & 3          & \(\mathcal{O}(n^2)\) \\
            Composite Link                & 4          & 3          & \(\mathcal{O}(n)\)   \\
            Interface (\ac{dsi}/\ac{dpi}) & 1          & 2          & \(\mathcal{O}(n^2)\) \\
            Data Component                & 2          & 3          & \(\mathcal{O}(n^2)\) \\
            Processing Component          & 4          & 3          & \(\mathcal{O}(n^2)\) \\
            Resource                      & 0          & 1          & \(\mathcal{O}(n^2)\) \\
        \end{tabular}}
    \caption{Additional predicates (F), numeric fluents (X), and
        grounded actions (A) required by additional object types.}\label{tab:object-complexity}
\end{table}

\begin{table}[t]
    \centering
    {\footnotesize
        \begin{tabular}{lccc}
            \textbf{Problem Size} & \textbf{F (AIBR)} & \textbf{X (AIBR)} & \textbf{A (AIBR)} \\
            \midrule
            2WFC, 1S              & 22 (8)            & 7 (6)             & 14 (6)            \\
            2WFC, \textbf{2S}     & 96 (24)           & 14 (12)           & 90 (22)           \\
            2WFC, \textbf{3S}     & 306 (39)          & 22 (19)           & 284 (51)          \\
            \textbf{4WFC}, 3S     & 843 (81)          & 28 (25)           & 1146 (109)        \\
        \end{tabular}}
    \caption{Grounding comparison: Predicates (F), Numeric Fluents (X), and
        Grounded Actions (A) before and after AIBR preprocessing.}\label{tab:grounding}
\end{table}

\subsection{Experiments Overview}

We possess two research questions we seek to answer about \worksworld{}:
\begin{enumerate}
    \item (RQ1) How does increasing the size of the workflow and resource graph
          affect the state space \ac{enhsp} searches?
    \item (RQ2) Does the domain definition permit scaling to large enough
          problems to plan and schedule workflows centrally across an enterprise?
\end{enumerate}

We design and execute two experiments. First, we leverage our configuration
and problem generator to observe state space growth
from Table~\ref{tab:object-complexity} by individually increasing object types.
Second, we explore a larger combination of components, interfaces, sites and
links that \ac{enhsp} can solve in reasonable time (one hour) and space (30~GB of
RAM).

\subsubsection{Environment and Configurations}

We use CloudLab~\cite{duplyakin2019cloudlab} with a reproducible
profile\footnote{https://gitlab.com/thpaul/k8s-promstack-profile} that
automatically configures our benchmarking environment. We deploy three
\emph{c6420}\footnote{https://docs.cloudlab.us/hardware.html} nodes (dual
sixteen-core processors at 2.6 GHz, 384GB DDR4 memory) running Kubernetes with
open-source tools for metrics collection, storage, and visualization. Container
images build and deploy to our Kubernetes cluster via a GitLab
pipeline\footnote{https://gitlab.com/thpaul/worksworld-benchmarks}.
We constrain solver containers to five cores and 30~GB RAM with a one-hour CPU
time limit per run. Each problem runs 20 independent times, and we report the
95th percentile with five-percent confidence intervals for measures with
variation.\ \ac{enhsp} runs with a 30GB Java heap using the
\(h_{\text{max}}^{\text{MRP}} + H\) heuristic~\cite{scala2020hmrp}
(\texttt{sat-hmrph}
flag\footnote{https://sites.google.com/view/enhsp/home/how-to-use-it}).

\section{Results}\label{sec:results}

\begin{figure}[t]
    \centering
    \includegraphics[width=\linewidth]{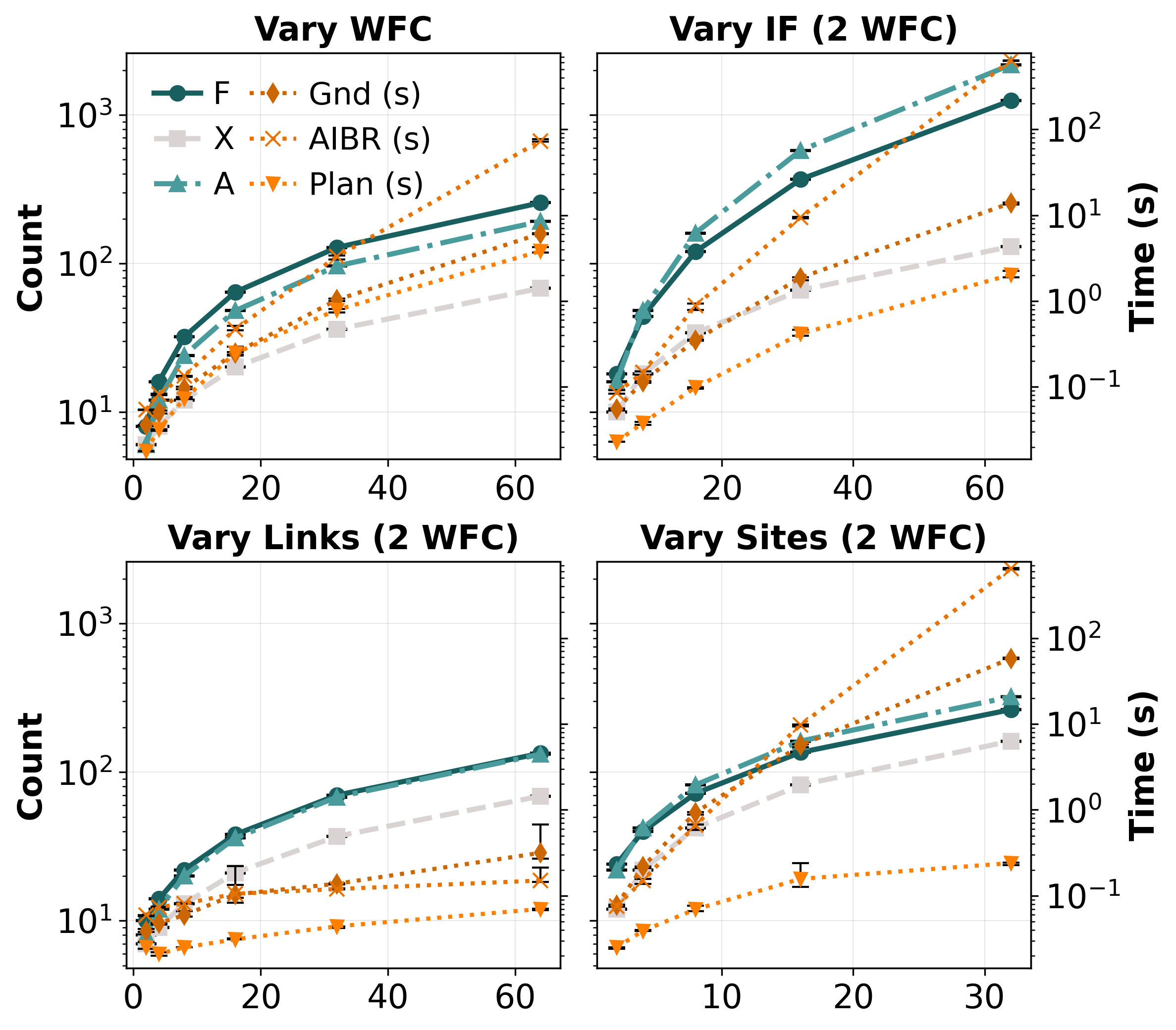}
    \caption{Planning and scheduling linear-chain workflows varying one
        aspect: workflow components, interfaces, direct links and
        sites.}\label{fig:individual-components}
\end{figure}

Figure~\ref{fig:individual-components} shows results from our first experiment
solving linear workflow chains while varying a single object type. As a
baseline, \(s_0\) contains one site with one intrasite link, one \ac{dsi} and
\ac{dpi}, and one scheduled source data component. For the top-left plot, we
exponentially increase workflow components to schedule from two until failure at
128. For interfaces, we fix two components (to schedule) while exponentially
increasing \acp{dsi}/\ac{dpi} from two each (4 total) to failure at 64 each (128
total). For direct links, we scale intrasite links exponentially to 64 without
failure. Finally, we add sites exponentially (each with one \ac{dsi}, one
\ac{dpi}, one intrasite link, and one direct link to a central site), succeeding
at 32 sites but failing at 64.

\begin{figure*}[t]
    \centering
    \includegraphics[width=\linewidth]{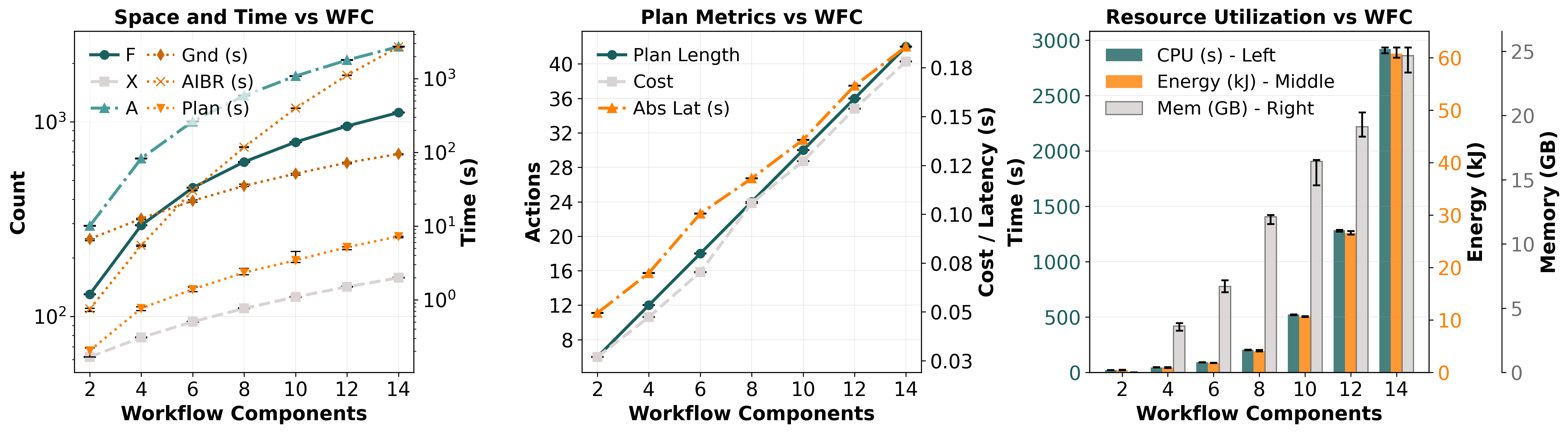}
    \caption{Planning and scheduling variable length linear-chain workflows on a complex
        resource graph (8S, 16IF, 15DL, 14CL).}\label{fig:complex-graph}
\end{figure*}

Figure~\ref{fig:complex-graph} presents our results from experiment two. From
our first experiment, we selected a resource graph that produces reasonable
planning times but is also representative of a mid-sized organization. We
include eight sites in our initial state, each with a \ac{dsi}/\ac{dpi} with
uniform \texttt{resource\_total} across sites but randomly varied
\texttt{resource\_available} values. We connect them via seven direct links
(with eight intrasite links) in a hub-spoke structure with the first site
connected to all other sites.  We then establish 14 composite links to permit
over half of the possible two-hop connections (\(\binom{7}{2} = 21\)). We
increase the size of the workflow scheduled by two's from two to 16 components;
14 components just meet our memory and time limits, 16 components fail. We
include total solver container energy measures from
Kepler~\cite{amaral2023kepler}, generated by reading the intel
RAPL\footnote{https://sustainable-computing.io/archive/design/kepler-energy-sources/}
metrics on each node, to report sustainability results and showcase the
extensibility of our Kubernetes benchmark environment.

\section{Discussion}\label{sec:discussion}

Figure~\ref{fig:individual-components} addresses RQ1. In the \emph{Vary WFC}
subplot, most metrics transition from exponential to sub-exponential growth
between 16 and 32 components, suggesting effective pruning, while \ac{aibr}
preprocessing remains exponential throughout. For direct links, \ac{enhsp}
handles added predicates and fluents efficiently. For interfaces, the planner
prunes effectively but preprocessing grows exponentially, becoming the planning
bottleneck. This effect intensifies when scaling sites (each adding two
interfaces and two links). In sum, \ac{aibr} preprocessing proves effective yet
emerges as the primary scalability bottleneck; accelerating this step is a
natural future research direction.

Figure~\ref{fig:complex-graph} addresses RQ2.\ \ac{enhsp} with \worksworld{} can
serve as a centralized planner for \emph{linear-chain} workflows at a realistic
resource graph size. The sub-exponential growth across all metrics, with the
exception of \ac{aibr} preprocessing, supports this conclusion. The resource
graph (8S, 16IF, 15DL, 14CL) represents a realistic mid-sized enterprise with
two cloud, two fog, and four edge sites. However, the linear-chain assumption
limits realism; we leave non-linear series-parallel workflows for future work.

\section{Related Works}\label{sec:related}

A broad body of literature addresses workflow scheduling and planning. We focus
on work most similar to our pipelined-workflow approach using \ac{ai} planning
techniques.

First, we discuss available planning domains for compute resource planning and
scheduling. Seipp~\etal{}~(\citeyear{seipp-et-al-zenodo2022}) provide an online
repository of existing domains. The \emph{Data-network} domain proves most
similar to \worksworld{}, modeling servers with different RAM capacities that
receive, process and send files across a network. Like \worksworld{}, goals
declare desired data formats on specific servers. However, this domain uses PDDL
2.2 level 1 (non-numeric planning) with PDDL 3.1 action costs, while
\worksworld{} fully supports numeric planning for more realistic resource
modeling. Numeric planning domains also exist from the International Planning
Competition numeric
track\footnote{https://github.com/ipc2023-numeric/ipc2023-dataset}, though none
closely resemble \worksworld{}.

Second, we discuss works leveraging \ac{ai} planning for similar problems.
Arshad~\etal{}~\citeyear{arshad2003deployment} develop Planit, which uses the
LPG planner~\cite{agnew1994planning} to dynamically reconfigure software systems
by scheduling components and point-to-point connectors, whereas we provide
\acp{dsi} for multi-component data sharing.
Kichkaylo~\etal{}~\citeyear{kichkaylo2003constrained} use node and link-level
abstractions to place components for applications (secure mail, webcast) by
compiling placement into planning problems.
Riabov~\etal{}~\citeyear{riabov2005planning} solve workflow planning by
declaring input and output streams but define a PDDL extension (Stream
Processing Planning Language) that prevents testing with domain-independent
planners. Shorabi~\etal{}~\citeyear{sohrabi2013htn} use \ac{htn} planning for
workflows in vendor-specific platforms, exploring optimal \emph{planning} of
graphs with 30--170 components; we explore satisficing planning and
\emph{scheduling} on resources in a vendor-agnostic manner. Most recently,
Amado~\etal{}~\citeyear{amado2025automated} group data pipeline tasks into
Kubernetes pods to minimize execution time for a vendor-specific application.
Compared to these works, we operate at a higher abstraction level, reasoning
over sites containing resources and generating plans in a vendor and
platform-agnostic manner.

\section{Conclusion}\label{sec:conclusion}

We introduced \worksworld{}, a numeric domain for integrated planning and
scheduling of permanent workflows. We showed empirically that a state-of-the-art
planner can solve the joint planning and scheduling problem we defined in
reasonable time and space for simple workflows on complex resource graphs. The
domain and our other engineering artifacts are available publicly to benefit the
AI planning community. Beyond future work mentioned earlier, we intend to
explore leveraging state-dependent action costs in the domain to permit greater
cost penalty when scheduling on high-utilization instances. This will require
normalizing work scheduled with our state-dependent \texttt{resource\_available}
fluent. Supporting state dependent action costs will also permit a more
conservative cost estimate based on actual rates instead of max rates for
workflow components. Second, we posit that increasing the types of \acp{dsi}
modelled in \worksworld{} could permit interesting tradeoffs for a planner. For
example, the planner may schedule latency-sensitive components on memory-based
interfaces at higher cost, and latency-tolerant components on disk-backed
interfaces at lower cost. Finally, our workflow and resource graph facilitate
easy division of work that may enable a hierarchy of planners to collaboratively
solve the scheduling portion of the problem. We envision a high-level planner
that produces the concrete workflow graph and delegates to each site's low-level
planner to schedule and connect its assigned components across available
interfaces. This approach may prove more efficient for problems with complex
resource and/or workflow graphs.

\bibliography{aaai2026}

\end{document}